# Title: Insulator-to-Metal Transition and Anomalously Slow Hot Carrier Cooling in a Photo-doped Mott Insulator

**Authors:** Usama Choudhry[1], Jin Zhang[2,3]*, Kewen Huang[4], Emma Low[4], Yujie Quan[1], Basamat Shaheen[1], Ryan Gnabasik[1], Jiaqiang Yan[5], Angel Rubio[2,6,7]*, Kenneth S. Burch[4], Bolin Liao[1]*

**Affiliations:**

[1]Department of Mechanical Engineering, University of California, Santa Barbara, California 93106, USA.

[2]Max Planck Institute for the Structure and Dynamics of Matter and Center for Free-Electron Laser Science, Luruper Chaussee 149, 22761 Hamburg, Germany.

[3]Laboratory of Theoretical and Computational Nanoscience, National Center for Nanoscience and Technology, Chinese Academy of Sciences, Beijing 100190, China.

[4]Department of Physics, Boston College, Chestnut Hill, Massachusetts 02467, USA.

[5]Materials Science and Technology Division, Oak Ridge National Laboratory, Oak Ridge, Tennessee 37831, USA

[6]Center for Computational Quantum Physics, The Flatiron Institute, 162 Fifth Avenue, New York, New York 10010, USA.

[7]Nano-Bio Spectroscopy Group, Universidad de País Vasco, 20018 San Sebastián, Spain.

*Corresponding authors. Email: jinzhang@nanoctr.cn, angel.rubio@mpsd.mpg.de, bliao@ucsb.edu

**Abstract:** Photo-doped Mott insulators can exhibit novel photocarrier transport and relaxation dynamics and non-equilibrium phases. However, time-resolved real-space imaging of these processes are still lacking. Here, we use scanning ultrafast electron microscopy (SUEM) to directly visualize the spatial-temporal evolution of photoexcited species in a spin-orbit assisted Mott insulator $\alpha$-$RuCl_3$. At low optical fluences, we observe extremely long hot photocarrier transport time over one nanosecond, almost an order of magnitude longer than any known values in conventional semiconductors. At higher optical fluences, we observe nonlinear features suggesting a photo-induced insulator-to-metal transition, which is unusual in a large-gap Mott insulator. Our results demonstrate the rich physics in a photo-doped Mott insulator that can be extracted from spatial-temporal imaging and showcase the capability of SUEM to sensitively probe photoexcitations in strongly correlated electron systems.

**One-Sentence Summary:** Extremely slow hot carrier cooling and a photo-induced insulator-to-metal transition are observed in a Mott insulator $\alpha$-$RuCl_3$ with scanning ultrafast electron microscopy.

**Main Text:** Understanding and tuning the complex coupling behaviors among charge, spin, orbital, and lattice degrees of freedom in strongly correlated electronic systems has been a central theme in condensed matter physics (*1*). As an important example, Mott insulators are insulating phases created by strong electron-electron repulsions in materials that should be metallic based on the non-interacting electron band theory (*2*). In a Mott insulator, the on-site Coulombic repulsion between electrons overcome the tendency of the electronic wavefunctions to spread and minimize their kinetic energy, leading to localized electrons and an insulating state. The close interplay between competing energy scales in Mott insulators can lead to collective behaviors causing large variations of physical properties in response to small perturbations. As a result, perturbing the ground state of a Mott insulator, e.g. by doping, changing temperature and pressure, applying external electrical and magnetic fields, and photoexcitation, can lead to a gamut of exotic phenomena of both fundamental interest and technological relevance, such as high-temperature superconductivity (*3*), metal-insulator transitions (*4*), and colossal magnetoresistance (*5*). Despite extensive research efforts, microscopic details of these processes remain to be understood.

Among various external perturbations, optical excitation provides a precise way to introduce mobile charges into Mott insulators and monitor their subsequent response with high temporal resolution enabled by ultrafast lasers (*6*). In addition, a large Mott gap can reduce trivial laser heating and suppress thermalization of hot photocarriers, which can give rise to novel nonequilibrium states (*6*). For these reasons, a range of optical pump-probe methods, including transient optical spectroscopy, time-resolved photoemission spectroscopy, and time-resolved X-ray spectroscopy, have been applied to investigate the dynamics of photoexcited carriers and, more interestingly, photo-induced nonequilibrium phases in Mott insulators (*7*, *8*). However, these existing studies based on optical spectroscopy and diffraction lack the spatial resolution to visualize the spatial-temporal evolution of photo-excited states, which, crucially, can reveal the impact of correlation on transient transport properties in Mott insulators.

Scanning ultrafast electron microscopy (SUEM) is an emerging technique that can image the spatial-temporal evolution of surface photocarrier dynamics. SUEM is a photon-pump-electron-probe technique that uses short electron pulses with picosecond duration to image the response of a sample surface after the impact of a photon pulse (*9*, *10*). It integrates the temporal resolution of femtosecond lasers with the spatial resolution of scanning electron microscopes (SEMs). The change of local secondary electron (SE) yield as a result of the optical excitation is measured and used to form contrast images (*11*). Given the shallow escape depth of SEs (a few nanometers), SUEM is highly sensitive to surface charge dynamics and has been used to study photocarrier diffusion in uniform semiconductors (*12*, *13*) and near interfaces (*9*, *14*). So far, SUEM has only been applied to image photocarrier dynamics in conventional semiconductors without strong electron correlation effects.

In this work, we report time-resolved imaging of photoexcited states in a van der Waals layered Mott insulator $\alpha$-$RuCl_3$ with SUEM. $\alpha$-$RuCl_3$ is a 4*d*-transition-metal halide composed of honeycomb layers made of adjoining $RuCl_6$ octahedra (*15*). Comprehensive photoemission, optical, magnetic, and transport measurements have suggested that it is a spin-orbit assisted Mott insulator with a wide Mott gap (*16*), which is a result of both strong electron correlations and spin-orbit coupling (*15*). The combination of a Mott insulating state and a honeycomb lattice of strong spin-orbit coupled transition metal ions has been identified as potential ingredients for actualizing Kitaev quantum spin liquids (*17*, *18*), leading to extensive studies of $\alpha$-$RuCl_3$ in pursuit of these states (*19*, *20*). In addition, theoretical studies have predicted that the Mott insulating state in $\alpha$-$RuCl_3$ can be sensitively tuned by photoexcitation, leading to optically driven magnetic (*21*, *22*) and insulator-to-metal (*23*) transitions. These rich physics make $\alpha$-$RuCl_3$ a particularly interesting material to study with SUEM. Here, we use SUEM to image the spatial-temporal evolution of photoexcited states in $\alpha$-$RuCl_3$ across different optical excitation fluences. At low optical fluences, we observe diffusion of photoexcited species with an extremely long hot carrier transport time beyond 1 ns at room temperature, which is almost one order of magnitude longer than known values in a  conventional semiconductor (*12*). At higher optical fluences, we observe nonlinear responses suggesting a photo-induced insulator-to-metal transition with an onset at as low as roughly 0.01 photoexcited electrons per unit cell. Our result suggests the electronic structure in $\alpha$-$RuCl_3$ can be sensitively controlled by a low concentration of photoexcited carriers at room temperature and showcases the capability of SUEM to probe spatial-temporal dynamics of photo-doped strongly correlated electronic systems.

Details of our SUEM setup is described elsewhere (*12*) and can be found in Supplementary Materials and Figs. S1 and S2. Briefly, we use an optical pump pulse [wavelength: 515 nm (2.4 eV), pulse duration: 150 fs, repetition rate: 5 MHz, beam diameter 30 $\mu$m] to excite the sample and use a delayed electron probe pulse ("primary electrons", or PEs; 25 electrons per pulse, energy: 30 keV). The electron pulses are generated by illuminating a Schottky electron gun with ultraviolet optical pulses (wavelength: 257 nm) from the same laser source. Upon impact of the PEs on the sample surface, SEs emitted from each location on the sample surface are collected by an Everhart-Thornley detector. The change of local SE yield as a result of the optical excitation, which is sensitive to the average electron energy, potential, and conductivity near the sample surface (*17*), is used to form SUEM contrast images. The spatial resolution of these images is on the order of the size of the PE beam (a few nanometers) and the time resolution is determined by the duration of the PE pulses (a few picoseconds with less than 100 electrons per pulse) (*12*, *24*). All experiments are conducted at room temperature with a vacuum level of $8\times10^{-7}$ torr.

$\alpha$-$RuCl_3$ crystals are exfoliated onto $Si/SiO_2$ substrates in a glovebox to minimize water exposure that can easily intercalate it (*25*). More details can be found in Supplementary Materials. Figure 1A shows representative SUEM images taken on a $\alpha$-$RuCl_3$ flake on an $Si/SiO_2$ substrate using an optical pump fluence of 30 $\mu$J/cm$^2$. Each contrast image is obtained by subtracting a reference image that is taken at a far negative time, and thus, the contrast in the images is representative of the change in SE emission from the sample surface because of photoexcitation. The lack of contrast at -60 ps indicates that the 200 ns interval between pump pulses is sufficiently long to allow the sample to equilibrate between pumping events. After time zero, a bright contrast emerges from the pump-illuminated region, indicating that SE emission from the area has increased. As established in SUEM, this contrast correlates with the distribution of carriers near the sample surface and results from the increased average energy of local electrons due to photoexcited carriers (*11*).

In the low fluence SUEM images shown in Fig. 1A, the evolution of the contrast suggests a diffusion process of a photoexcited species on the time scale of hundreds of picoseconds to a few nanoseconds. The diffusion process can be quantified by fitting the SUEM contrast to a two-dimensional Gaussian function with time-dependent magnitude and radius (*12*). The fitted time-dependent magnitude is shown in Fig. 1B, suggesting a recombination lifetime of $2.3\pm0.3$ ns. The fitted time-dependent radius is shown in Fig. 1C. The diffusivity of the photocarriers depends on time, due to the "hot" state of the photocarriers immediately after excitation since the pump energy (2.4 eV) is much higher than the optical gap of 1 eV in $\alpha$-$RuCl_3$ (*15*). The diffusivity decays as the hot photocarriers thermalize and cool down to the band edges through interactions with phonons and spins. Hot photocarrier transport has been observed with SUEM in a range of conventional semiconductors (*10*, *12*, *26*), which can persist for over 200 ps in boron arsenide due to a hot phonon bottleneck effect (*12*). The hot photocarrier diffusion behavior can be quantified by analyzing the SUEM contrast using a diffusion equation with an exponentially decaying diffusivity with a time constant $\tau$, $D(t) = \left(D_i - D_0\right) e^{-t/\tau} + D_0$, where $D_i$ is the effective diffusivity immediately after photoexcitation, and $D_0$ is the equilibrium diffusivity at 300 K. The fitting is shown in Fig. 1C for an optical fluence of 30 and 40 $\mu$J/cm$^2$, where an extremely long hot carrier transport time $\tau$ of roughly 1 ns at 30 $\mu$J/cm$^2$ (1.5 ns at 40 $\mu$J/cm$^2$) and an initial diffusivity of 2,000 cm$^2$/s are extracted. The initial diffusivity is orders of magnitude lower than that in silicon (*26*). Our low fluence result suggests that the hot photocarriers in $\alpha$-$RuCl_3$ diffuse more slowly than those in conventional semiconductors but cool down to the band edges on a much longer time scale (at least 7 times longer than that in boron arsenide).

We observe qualitatively different SUEM contrasts at increased optical pump fluences. Figure 2A shows representative SUEM contrast images using a fluence of 400 $\mu$J/cm$^2$. This fluence corresponds to roughly 0.02 photoexcited electrons per unit cell near the center of the Gaussian pulse (see Supplementary Materials for more discussions). No contrast is visible at negative times, once again indicating that the sample is adequately relaxing between pump events. After a few ps, instead of observing a Gaussian region of bright contrast, a ring of bright contrast is present with a central area of dark contrast. The area of bright contrast expands slightly over a few hundred picoseconds before relaxing on a timescale roughly similar to the low fluence measurements. However, the region of dark contrast persists for the full 4 nanoseconds that are accessible to the experiment. Additional SUEM image series are shown in the Supplementary Materials for intermediate fluences between 10 and 400 $\mu$J/cm$^2$.

While there is some deviation in the observed image contrast due to variation in absorption between different locations on the different samples, two general observations can be made: (1) the overall contrast magnitude is observed to decrease with increasing pump fluence, which is in sharp contrast to all previous SUEM studies, and (2) the magnitude of the dip in the center of the pump-illuminated region increases with fluence. Therefore, at intermediate fluences between 30 and 400 $\mu$J/cm$^2$, this feature begins to manifest first as a small decrease in contrast while remaining positive in magnitude. At higher fluences, the center region becomes dark, indicating that SE emission is suppressed in the region of the sample experiencing the greatest instantaneous fluence. These features can be seen clearly in Fig. 2B and C, where line cuts of the SUEM contrasts at different optical fluences are shown. A systematic examination of the SUEM images taken at different optical fluences is shown in Fig. S3, which indicates an onset of the nonlinear feature at a fluence around 40 $\mu$J/cm$^2$, although variation of this critical fluence between different locations of different samples is observed. We also carefully design the order of measurements with different optical fluences to rule out the potential trivial explanation that the observed contrast is due to some laser-induced damage of the samples (see Fig. S4).

Figure 3A and B summarize the evolution of the magnitude of the SUEM contrast in different regions of interest spanning fluences from 50 to 400 $\mu$J/cm$^2$. Fig. 3A shows the change in contrast for the center region, while Fig. 3B tracks the brightest region on the shoulder of the pump spot. As observed, the signal level in the center of the pump spot is lower than that on the edge, and the overall signal level decreases as the pumping fluence increases. Figures 3C and D show the normalized change in contrast in order to better compare changes in relaxation at different fluences. Figure 3C indicates that there is no obvious change in relaxation dynamics in the center region at low fluences, but at high fluences, the behavior becomes obviously different. At 250 $\mu$J/cm$^2$, the contrast in the center decays quicker than at lower fluences, while at 300 and 400 $\mu$J/cm$^2$, the contrast becomes persistently negative. The transition starts to occur around 200 $\mu$J/cm$^2$, which corresponds to roughly 0.01 electrons photoexcited per unit cell near the center of the pump beam. Curiously, the contrast on the bright edge of the pump spot (Fig. 3D) does not show any change in relaxation dynamics as a function of fluence, with a recombination lifetime of roughly 2.5 ns across all fluences.

In Mott insulators, excited carriers can be described as negative doublons, corresponding to doubly occupied sites, and positive holons, corresponding to unoccupied sites. There are theoretical predictions that the relaxation time scales for doublons and holons in Mott insulators increases rapidly when the onsite Coulomb energy *U* becomes much larger than the bandwidth due to a thermalization bottleneck (*27*, *28*), which can potentially explain the long relaxation lifetime we observe at low optical fluences. However, the extremely long hot photocarrier transport time over 1 ns has not been observed in a Mott insulator. Doublons and holons can also form Mott-Hubbard excitons (MHEs), which were proposed to explain time-resolved two-photon photoemission spectroscopy and transient reflection spectroscopy measurements on $\alpha$-$RuCl_3$ (*29*). The long hot photocarrier transport time observed in our experiment is similar to a slow timescale measured in their experiment (*29*), which was interpreted as the MHE lifetime. A potential explanation for the persistent hot photocarrier transport can be a cooling bottleneck between bands with pseudospin $J_{eff} = 1/2$ and $J_{eff} = 3/2$ in $\alpha$-$RuCl_3$ (*30*, *31*), which is a unique feature in spin-orbit assisted Mott insulators.

The high-fluence SUEM features can be explained by a photo-induced insulator-to-metal transition (*32*). SE yield sensitively depends on the local electrical conductivity (*33*), thus making SUEM a suitable tool to detect significant photo-induced conductivity changes. In an insulating material, slight charging of the sample due to primary electron injection can lead to surface electrostatic fields that enhance the SE yield. A photo-induced metallic state will suppress this electrostatic field and lead to reduced SE yield. This effect can explain the reduced overall SUEM contrast at increased optical fluences, which is distinct from previous SUEM experiments. Furthermore, insulators typically have higher SE yield due to reduced electron-electron scatterings experienced by the SEs as they migrate through the sample before escaping from the surface (*33*). Thus, the photo-induced metallic conductivity can reduce local SE yield, which competes with the more common effect that an increased average local electron energy due to photoexcitation enhances SE yield. This trend is summarized in Fig. 4. In the center of the Gaussian optical pump beam, the excitation level is the highest that leads to the most suppressed SE yield, which manifests as a central dip or dark region in the SUEM images.

In a previous work (*23*), we have theoretically explored the possibility of a photo-induced insulator-to-metal transition in $\alpha$-$RuCl_3$ using *ab initio* time-dependent density functional theory (TDDFT) with the ACBN0 functional developed to simulate electronic properties of strongly correlated materials (*34*). We found that above-gap photoexcitation significantly reduces the onsite energy $U$, which can lead to melting of the Mott gap and an insulator-to-metal transition in $\alpha$-$RuCl_3$ on a subpicosecond timescale. However, we caution that it is currently computationally infeasible to match the experimental time scale using TDDFT and, thus, a direct comparison between the TDDFT simulation in (*23*) and our experimental observation in this work cannot be made reliably. Nevertheless, the TDDFT study points to photo-induced band structure renormalization as a possible mechanism for our observed insulator-to-metal transition. While photo-induced metallic states have been observed in small-gap Mott insulators (*35–37*), its realization in a large-gap Mott insulator such as $\alpha$-$RuCl_3$ at a low excitation level is quite unusual and more studies are required to clarify its nature.

In summary, our SUEM images in $\alpha$-$RuCl_3$ suggest an anomalously long hot photocarrier transport time at low optical fluences, likely due to a cooling bottleneck between bands with different pseudospins, and a photo-induced insulator-to-metal transition at higher optical fluences. Both findings indicate novel transport and relaxation mechanisms in photo-doped Mott insulators. These results also demonstrate SUEM as a sensitive probe to photo-induced nonequilibrium phases in strongly correlated electron systems.

**Acknowledgments:**

**Funding:** We thank Yuki Motome and Mengkun Liu for helpful discussions. This work is based on research supported by US Air Force Office of Scientific Research under the award number FA9550-22-1-0468 (for the study of hot photocarrier dynamics) and by the US Army Research Office under the award number W911NF2310188 (for the development of SUEM). JY was supported by the U.S. Department of Energy, Office of Science, National Quantum Information Science Research Centers, Quantum Science Center (for sample growth).

**Author contributions:**

Conceptualization: KB, BL

Sample preparation: JY, KH, KB

SUEM development: UC, YQ, BS, RG

SUEM measurement: UC

SUEM data analysis: UC, BL

Theoretical analysis: JZ, AR

Funding and project administration: KB, BL

Supervision: AR, KB, BL

Writing – original draft: UC, BL

Writing – review & editing: all authors

**Data and materials availability:** All data are available in the main text or the supplementary materials.

**Supplementary Materials**

Materials and Methods

Supplementary Text

Figs. S1 to S11

Supplementary Movies

References (*37–38*)

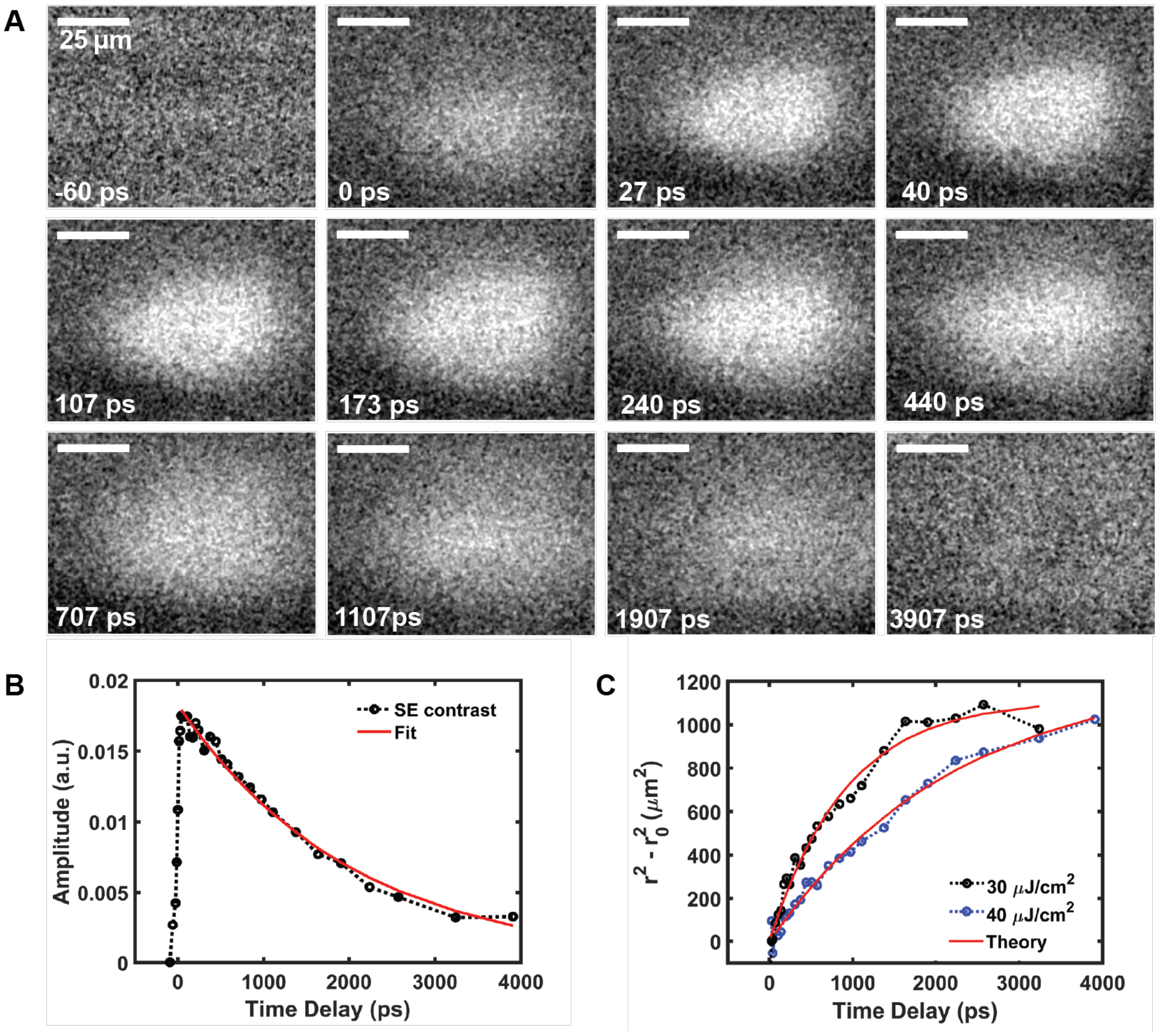

**Fig. 1. SUEM contrast images of $\alpha$-RuCl$_3$ taken at low optical fluences.** (**A**) SUEM contrast images taken as a function of delay time with an optical pump fluence of 30 $\mu$J/cm$^2$. The bright contrast indicates enhanced secondary electron yield in photo-excited region. (**B**) Amplitude of the SUEM contrast (fitted to two-dimensional Gaussian functions) as a function of delay time with an optical pump fluence of 30 $\mu$J/cm$^2$, suggesting a photocarrier recombination lifetime of 2.2 ns. (**C**) Radius of the SUEM contrast (fitted to two-dimensional Gaussian functions; original radius $r_0$ subtracted) as a function of delay time at optical fluences of 30 and 40 $\mu$J/cm$^2$. The theoretical model used to fit the data is explained in Supplementary Materials.

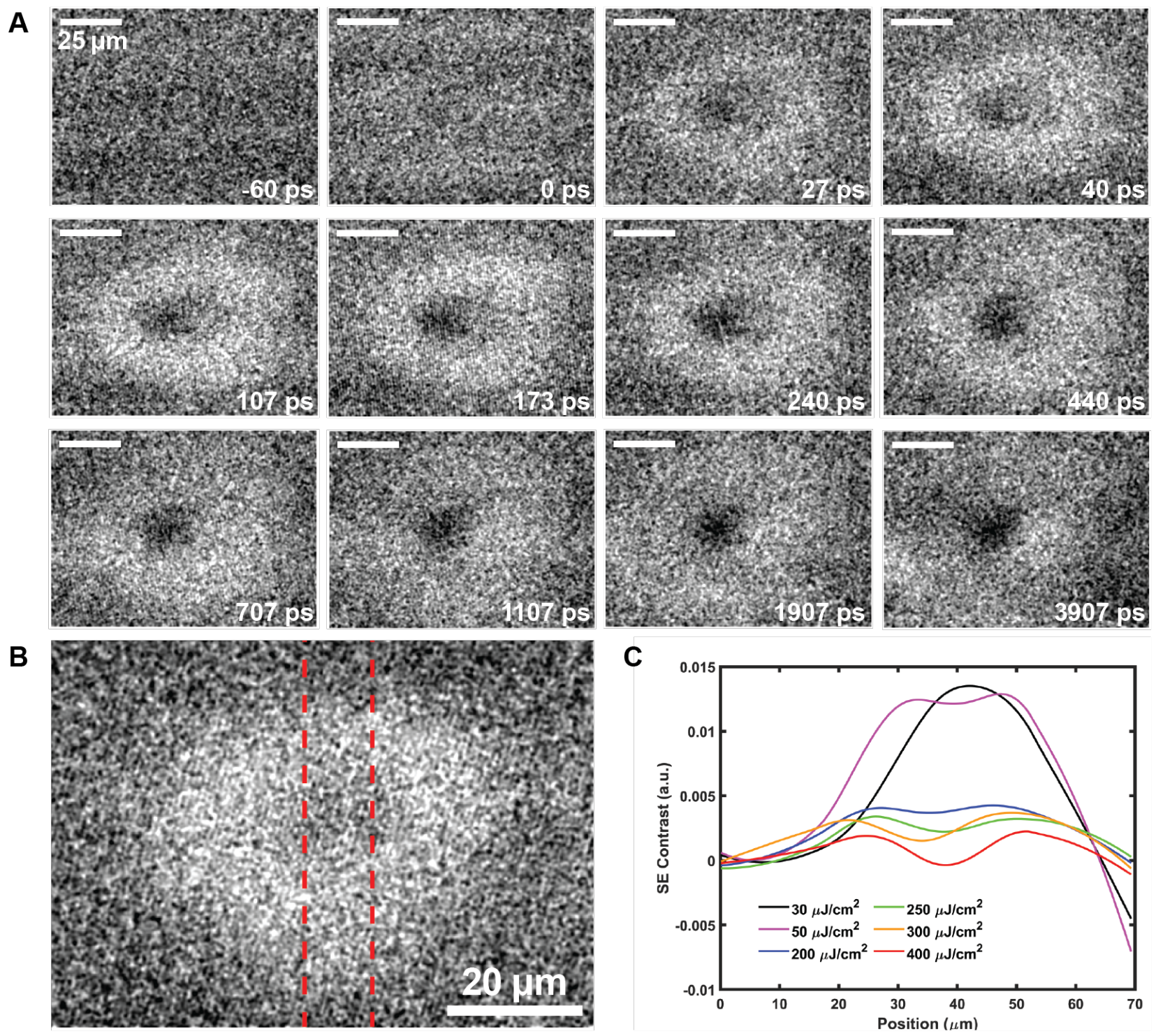

**Fig. 2. SUEM contrast images of $\alpha$-RuCl₃ taken at high optical fluences.** (**A**) SUEM contrast images taken as a function of delay time with an optical pump fluence of 400 $\mu$J/cm$^2$. Highly nonlinear features including a center dark region and a bright outer region emerge at higher optical fluences. (**B**) Representative SUEM image with red dashed lines indicating the regions of interest, where the linecuts shown in (**C**) are taken. (**C**) Linecuts of the SUEM contrast magnitude at different optical fluences, showing a reduced contrast in the center and overall suppressed contrast as the optical pump fluence is increased.

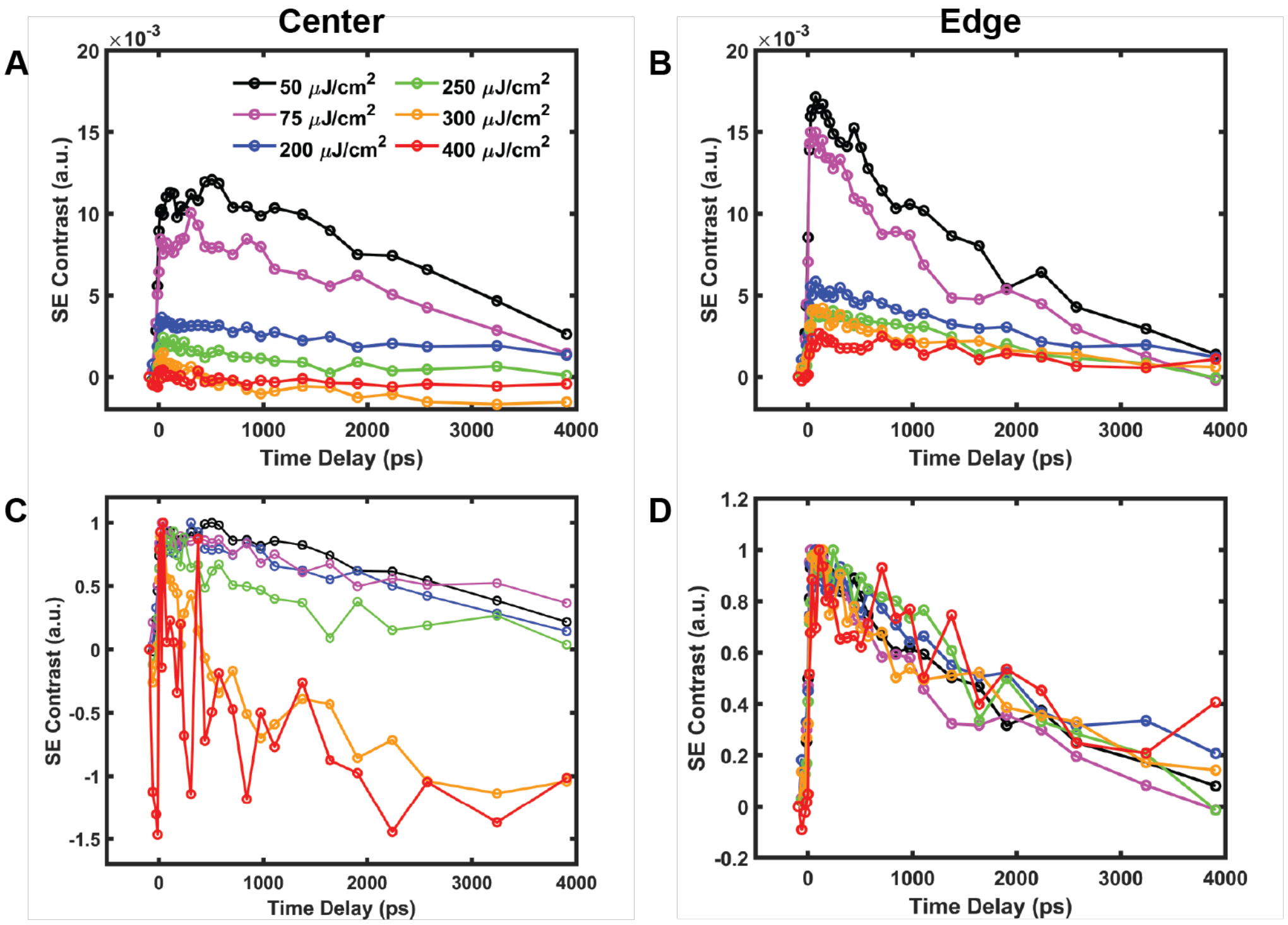

**Fig. 3. Evolution of the SUEM contrasts at high optical fluences.** The evolution of the SUEM contrast in the center dark region and the outer bright region at different optical pump fluences is shown in (**A**) and (**B**). Normalized versions are shown in (**C**) and (**D**). While the relaxation dynamics in the outer bright region is independent of the optical pump fluence, a sharp transition near 250 $\mu$J/cm$^2$ is observed in the center region, signaling a photo-induced insulator-to-metal transition. Significant noise at higher optical fluences is due to reduced overall SUEM contrast.

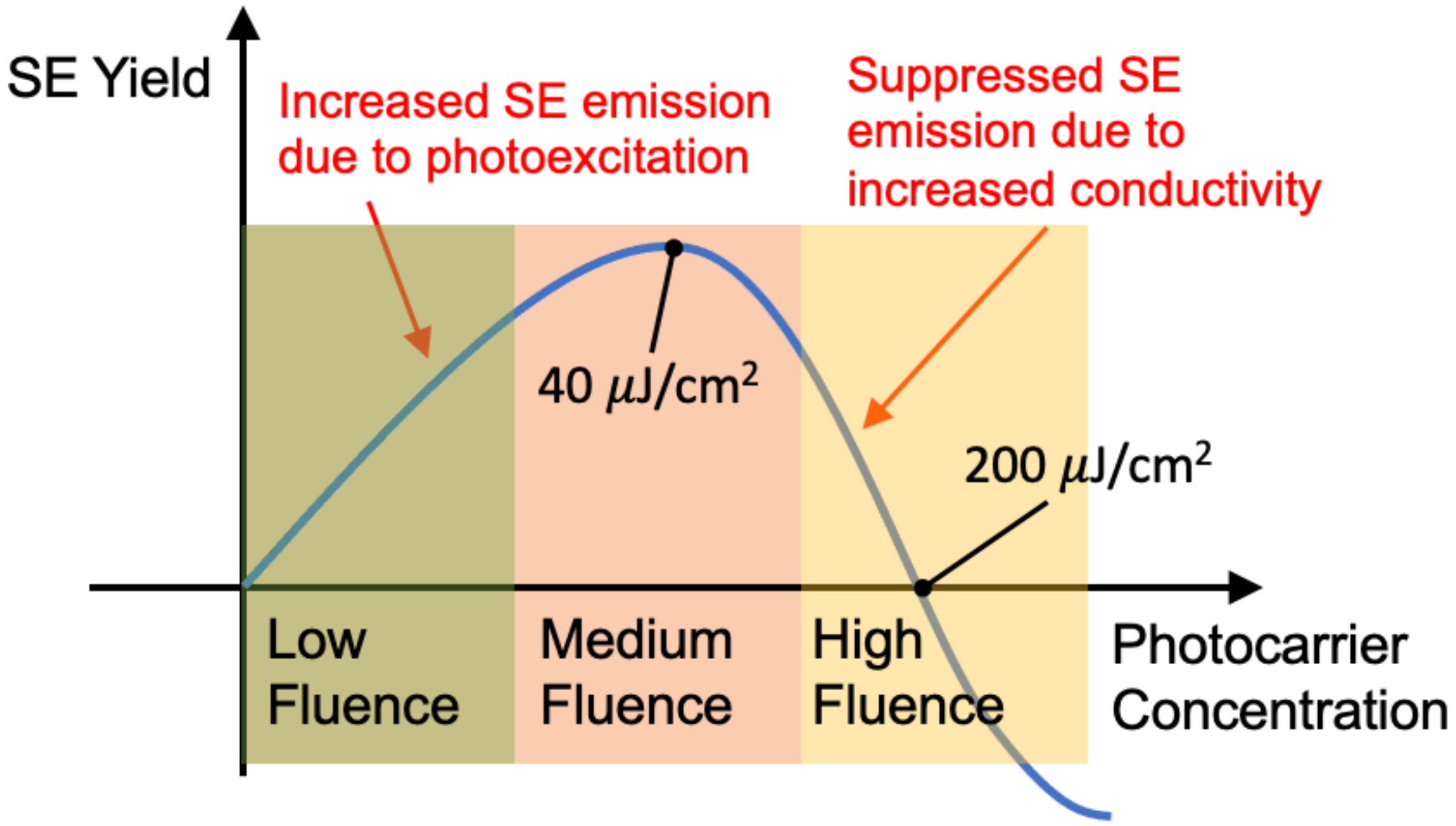

**Fig. 4. Physical picture of photo-induced SUEM contrast in $\alpha$-$RuCl_3$.** Schematic showing the competing mechanisms affecting the secondary electron (SE) yield as a function of photocarrier concentration. At low optical fluences, increased average electron energy due to photoexcitation enhances the SE yield. At higher optical fluences, increased local conductivity due to photo-induced insulator-to-metal transition suppresses the SE yield.

# Supplementary Materials for

## Insulator-to-Metal Transition and Anomalously Slow Hot Carrier Cooling in a Photo-doped Mott Insulator

Usama Choudhry, Jin Zhang, Kewen Huang, Emma Low, Yujie Quan, Basamat Shaheen, Ryan Gnabasik, Jiaqiang Yan, Angel Rubio, Kenneth S. Burch, Bolin Liao

Correspondence to: jinzhang@nanoctr.cn, angel.rubio@mpsd.mpg.de, bliao@ucsb.edu

**This PDF file includes:**

Materials and Methods
Supplementary Text
Figs. S1 to S11

## Materials and Methods

Scanning ultrafast electron microscopy

This section provides detailed description on the SUEM setup employed in this work (Fig. S1 and Fig. S2). A fundamental infrared (IR) laser pulse train (Clark MXR IMPULSE, wavelength: 1030 nm, pulse duration: 150 fs, repetition rate: 5 MHz) is directed to frequency-doubling crystals to create the visible pump beam (wavelength: 515 nm with tunable power) and the ultraviolet (UV) photoelectron excitation beam (wavelength: 257 nm, power: 15 mW). The visible pump beam travels variable distances adjusted by a mechanical delay stage (Newport DL600, delay time range: -0.7 ns to 3.3 ns). The UV excitation beam is directed through a transparent window on the column of an SEM (ThermoFisher Quanta 650 FEG) and onto the apex of a cooled Schottky field emission gun (a zirconium-oxide-coated tungsten tip), generating electron pulses with sub-picosecond durations via the photoelectric effect. An electron current (I) of 20 pA is used in this experiment, corresponding to 25 electrons per pulse. A time resolution of 2 ps is expected for 25 electrons per pulse. The photo-generated electron pulses are accelerated inside the SEM column to 30 keV kinetic energy, and are finely focused to nanometer size through the electron optics in the SEM. The beam dwell time at each pixel was 300 ns. Each SUEM contrast image at a given delay time represented an average of 2000 to 4000 images. During the measurements, the photocathode was refreshed every 60 minutes to prevent the fluctuation of the cathode work function. A mechanical coupling system is built to make a rigid connection between the SEM air-suspension system and the optical table hosting the laser and the optical system to minimize the relative vibration that affects the alignment at the photocathode.

Sample preparation

$\alpha$-$RuCl_3$ single crystals used in this study were grown by a self-selecting vapor transport technique reported elsewhere (*1*). $\alpha$-$RuCl_3$ flakes were then exfoliated from a bulk single crystal using Nitto tape. Flakes used in this study have a thickness ranging from 1 to 3 $\mu$m, which are much larger than the optical penetration depth to avoid the interference from the substrate. These flakes were exfoliated onto n-doped Si/$SiO_2$ chips that were first cleaned by sonication in acetone for 10 minutes, then rinsed with IPA, and then $dH_2O$. These chips had been treated previously with photoresist in order to create markers on their surface for easy location in the SEM. The photoresist was removed during this cleaning process. The water was blown off with an argon gas gun, and the chip was baked for 5 minutes at 105 °C to remove any residual water on the surface. The samples were then searched using a microscope, and the surfaces of each flake were examined using dark field microscopy for a 1-second exposure time to see if there were any visible debris or defects on the flake's surface. The flakes were then characterized by Raman spectroscopy using 200 μW laser power. The single spectra were taken with 1800 gr/mm for 180. All studied flakes revealed Raman spectra consistent with bulk.

**Supplementary Text**

Diffusion model

We used a simple diffusion model to analyze our experimental data. Given the axial symmetry of our experimental geometry, the diffusion process of photocarriers is governed by the following diffusion equation:

$$\frac{\partial n}{\partial t} = D(t)\left[\frac{1}{r}\frac{\partial}{\partial r}\left(r\frac{\partial n}{\partial r}\right)\right], \quad \text{(S-1)}$$

where $n$ is the density of photocarriers, $D(t)$ is a time-dependent effective diffusivity, $r$ is the radial distance from the center of the optically excited area, $z$ is the depth into the sample from the surface. We assumed the effective diffusivity $D(t)$ decays exponentially in time as the hot carriers cool down, as explained in the main text. Immediately after the photoexcitation, the distribution of the photocarrier density has the following form:

$$n(r, t = 0) = n_0 e^{-2r^2/R_0^2}, \quad \text{(S-2)}$$

where $n_0$ is the photocarrier density at the center of the illuminated area on the surface and $R_0$ is the $1/e^2$ radius of the optical pump beam. By assuming that the surface photocarrier density distribution follows a Gaussian distribution with a time-dependent radius: $n(r,t) = n_0(t)\, e^{-2r^2/R^2(t)}$, substitution of this solution into the radial diffusion equation leads to the following equation governing the time dependence of $R(t)$:

$$\frac{dR}{dt} = \frac{2D(t)}{R} = \frac{2}{R}\left[\left(D_i - D_0\right)e^{-t/\tau} + D_0\right]. \quad \text{(S-3)}$$

This equation can be analytically solved to give the following time dependence of $R(t)$:

$$R^2 - R_0^2 = 4\left(D_i - D_0\right)\tau\left(1 - e^{-t/\tau}\right) + 4D_0 t. \quad \text{(S-4)}$$

We use eqn. (S-4) to fit our experimental data at low optical fluences to extract the hot photocarrier transport time. Good agreement between this model and our experimental data is achieved.

Estimation of photocarrier concentration

The optical absorption depth at 2.4 eV in $\alpha$-$RuCl_3$ is roughly 318 nm (*2*). The optical reflectance at 2.4 eV of our sample is estimated to be 15% to 20%. Based on these optical properties, the average photocarrier concentration excited by a certain optical fluence can be estimated by

$$n = (1 - R)\frac{F}{E_{ph} d}, \quad \text{(S-5)}$$

where $R$ is the optical reflectance, $F$ is the optical fluence, $E_{ph}$ is the photon energy, and $d$ is the optical penetration depth. For a Gaussian beam, the peak fluence at the center is two times the average fluence. Therefore, the peak photocarrier concentration at the center of the beam is two times the average concentration estimated by Eqn. S-5.

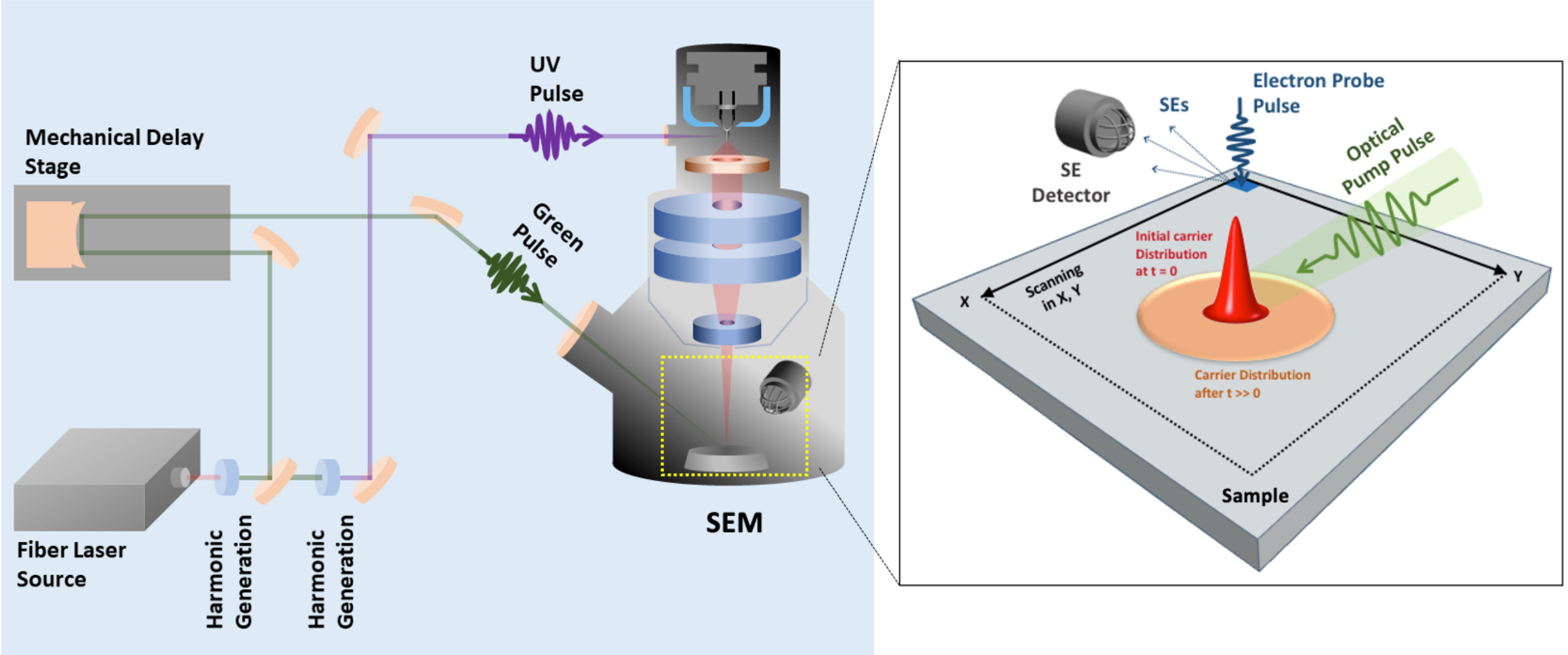

**Fig. S1. Schematic of the SUEM setup.** Left: an SEM is coupled to an ultrafast laser source via real-space optical coupling. The green optical beam from the laser is sent to the sample chamber to serve as the optical pump, and the ultraviolet beam is sent to the electron gun to generate short electron pulses as the probe. Right: secondary electron (SE) contrast imaging mechanisms in SUEM.

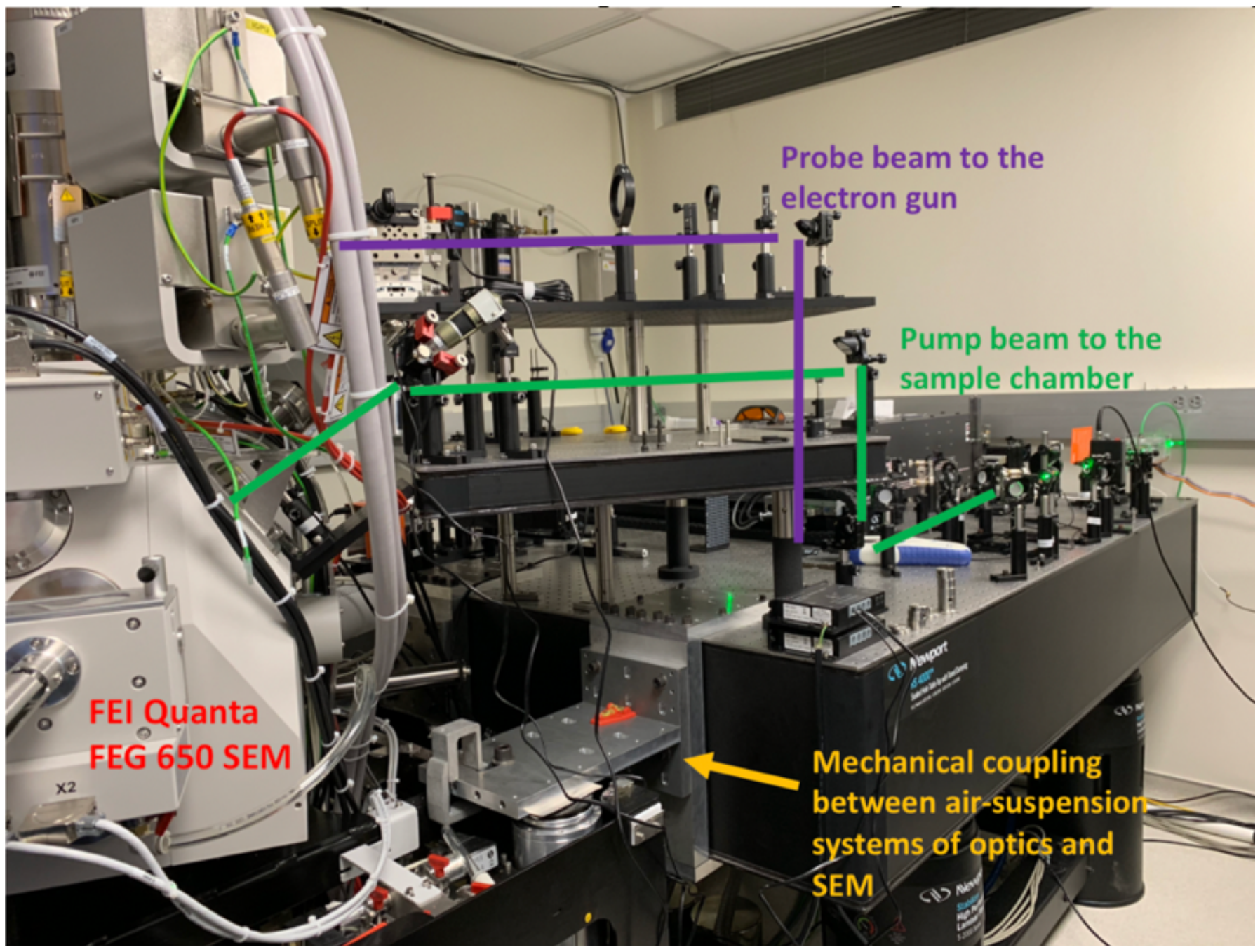

**Fig. S2. Picture of the SUEM setup at UCSB**.

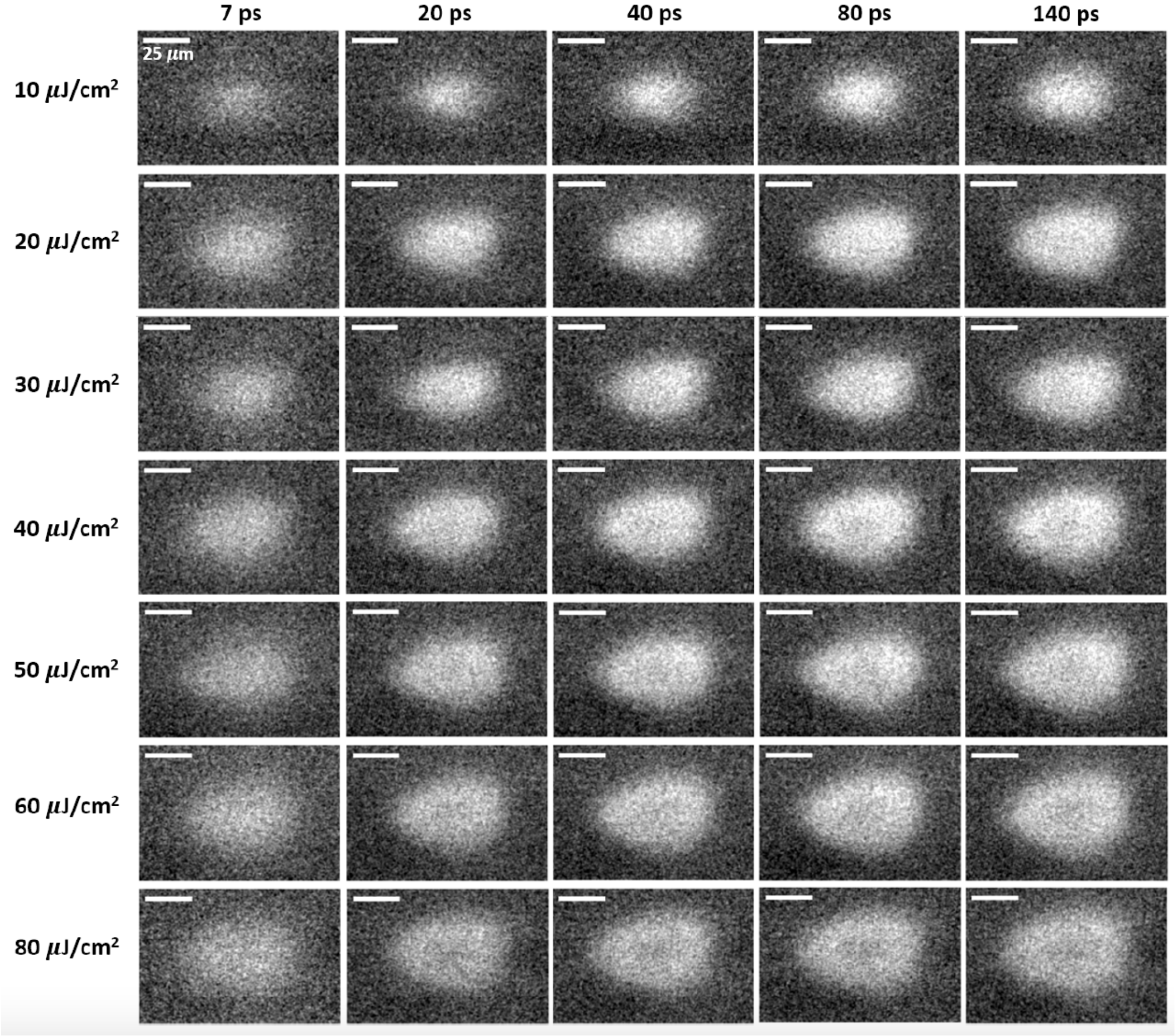

**Fig. S3. Observing critical crossover fluence on $\alpha$-$RuCl_3$.** SUEM contrast images taken on $\alpha$-$RuCl_3$ flake at several different fluences for a few selected time points. The nonlinear response at the center of the pump-illuminated area emerges at fluences greater than 40 $\mu$J/cm$^2$. The presence of non-Gaussian contrast indicates that the sample's response to photoexcitation is nonlinear.

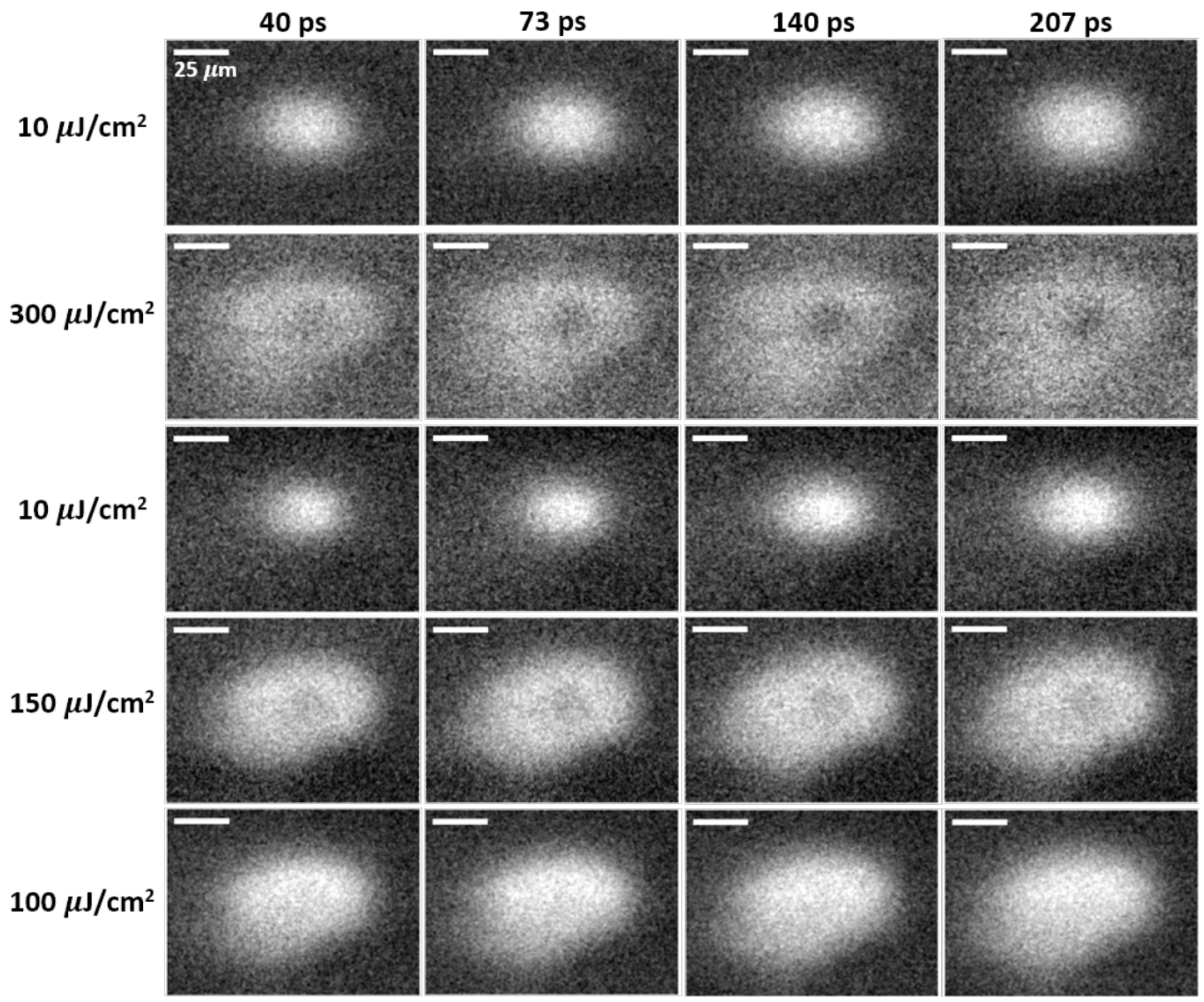

**Fig. S4. Sample damage test on $\alpha$-RuCl$_3$.** SUEM contrast images taken on $\alpha$-RuCl$_3$ at several different fluences for a few selected time points. The measurements were performed in the order they are shown, and on the same spot on the sample. The linear response is recovered after each subsequent high fluence measurement, showing that the sample is not being damaged.

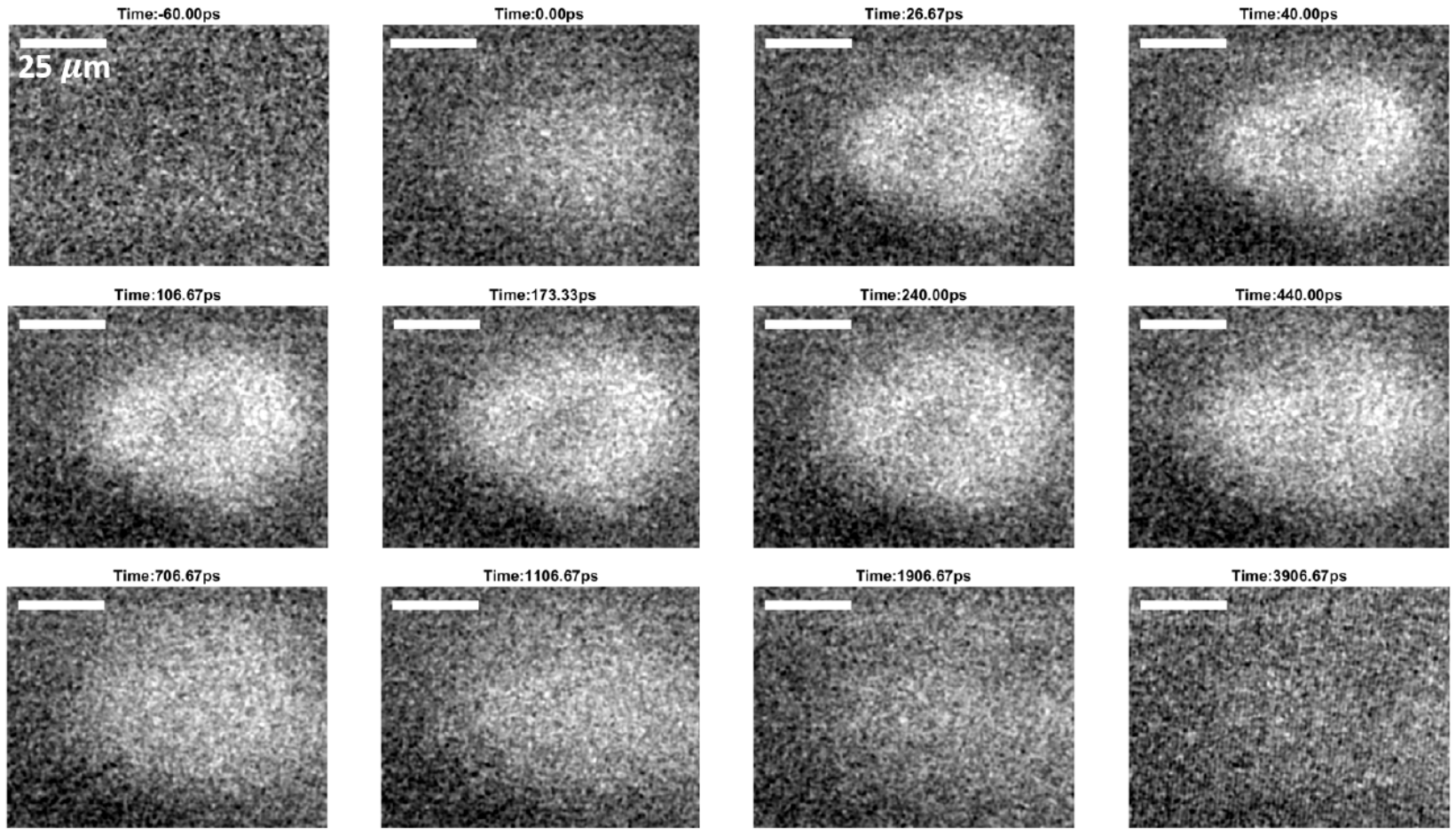

**Fig. S5. Additional SUEM contrast images taken on $\alpha$-$RuCl_3$.** Optical fluence: 50 μJ/cm$^2$.

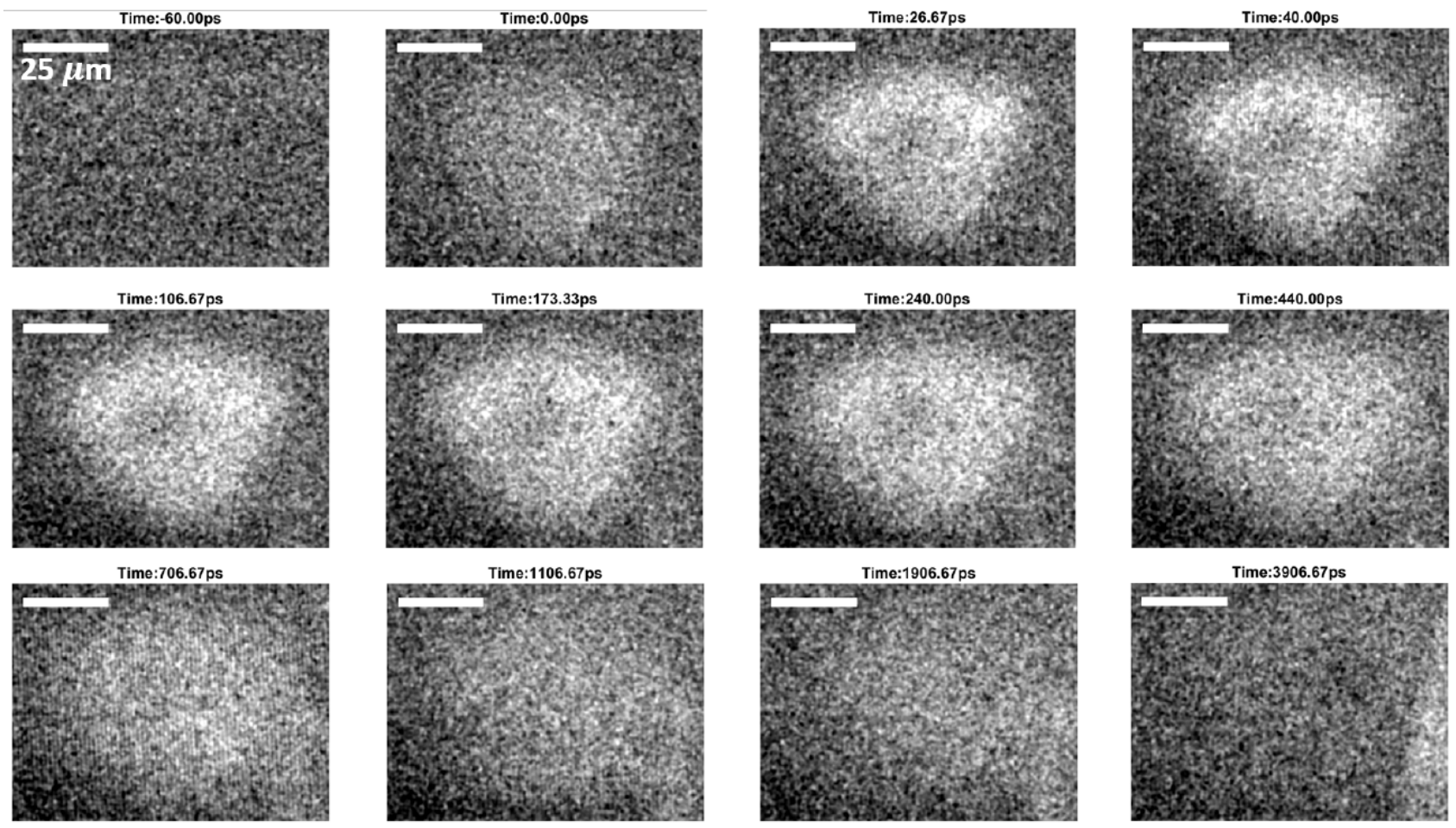

**Fig. S6. Additional SUEM contrast images taken on $\alpha$-$RuCl_3$.** Optical fluence: 75 μJ/cm$^2$.

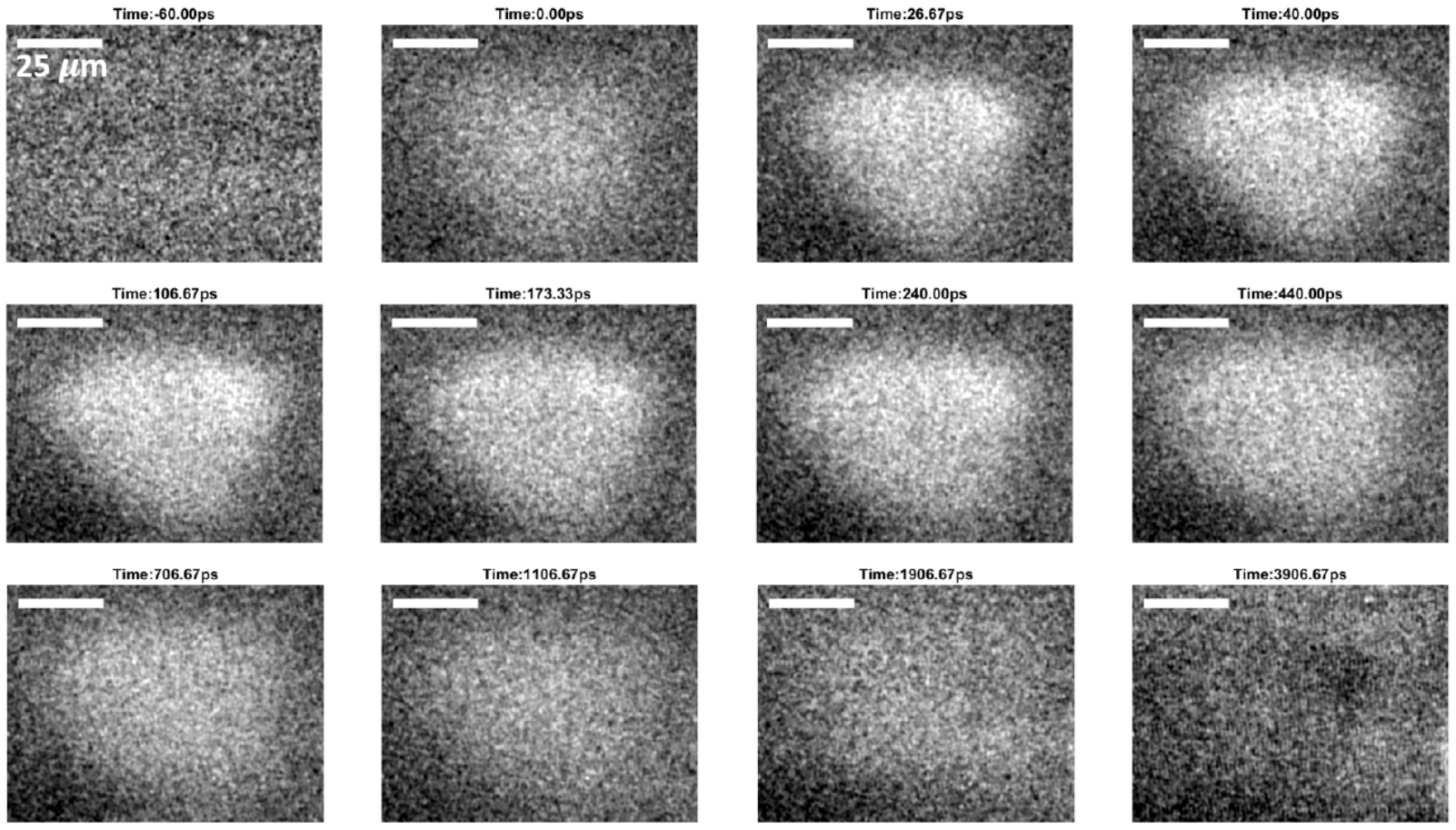

**Fig. S7.** Additional SUEM contrast images taken on $\alpha$-$RuCl_3$**.** Optical fluence: 100 μJ/$cm^2$.

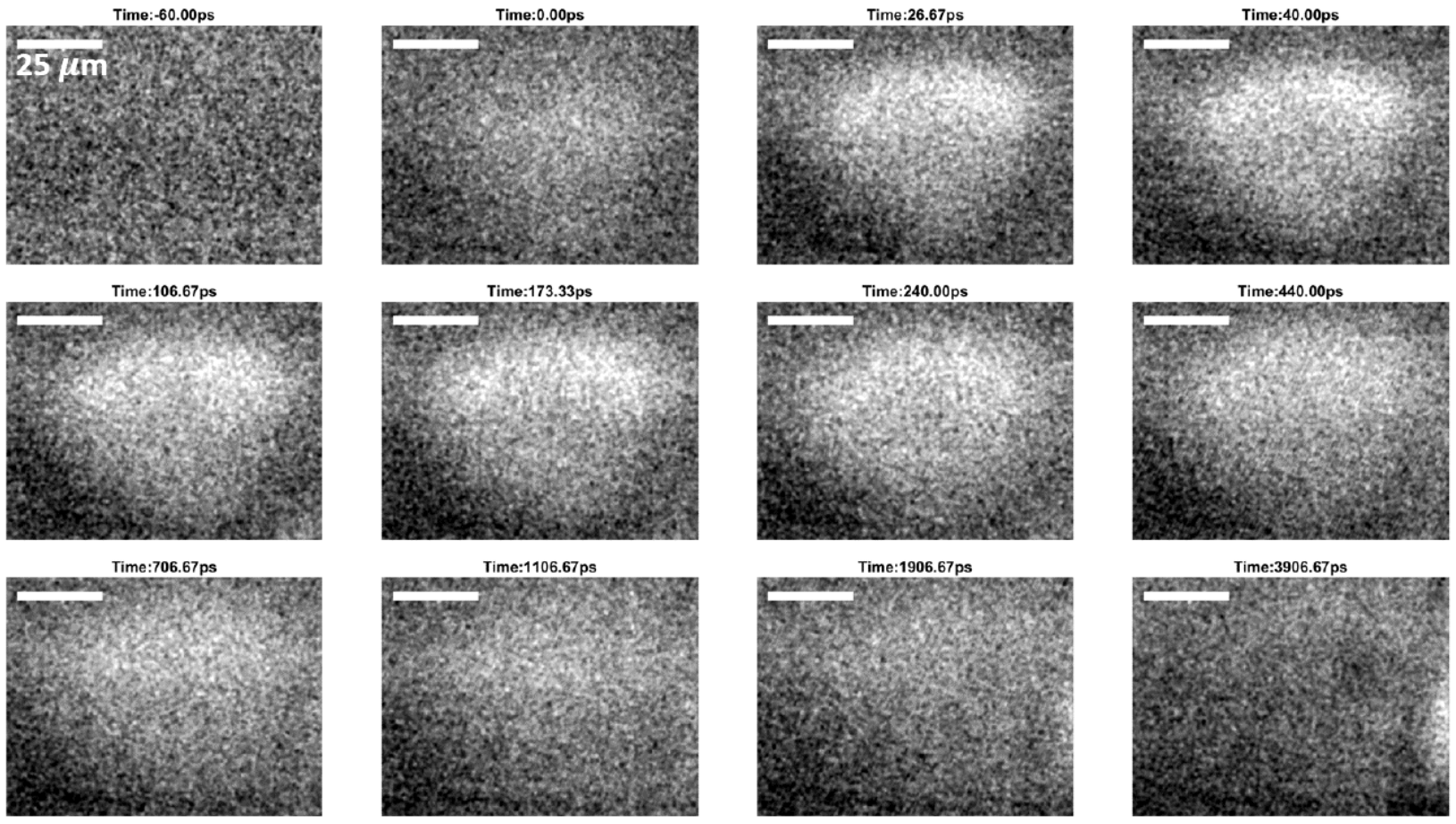

**Fig. S8.** Additional SUEM contrast images taken on $\alpha$-$RuCl_3$**.** Optical fluence: 150 μJ/cm$^2$.

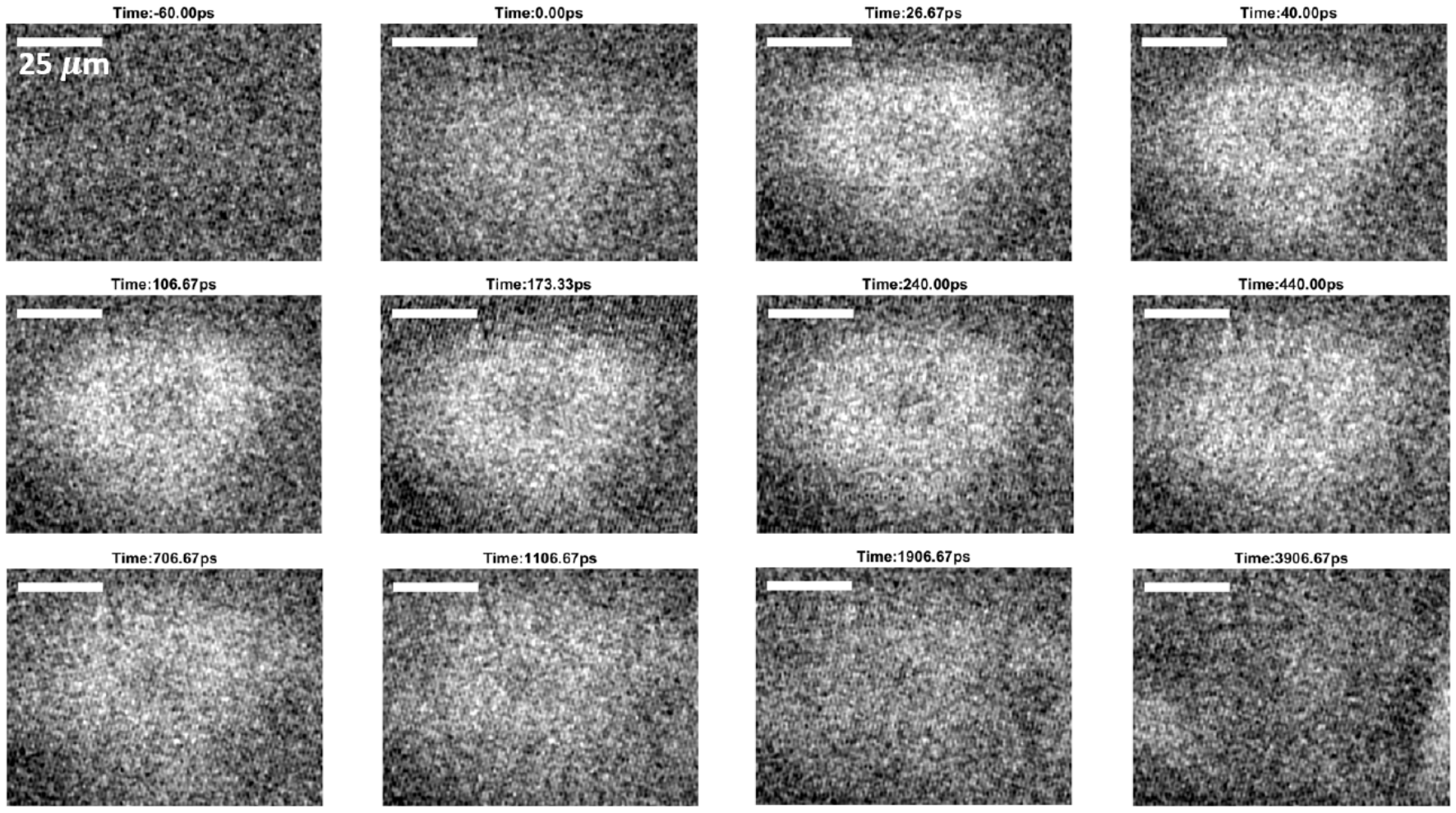

**Fig. S9.** Additional SUEM contrast images taken on $\alpha$-$RuCl_3$**.** Optical fluence: 200 μJ/$cm^2$.

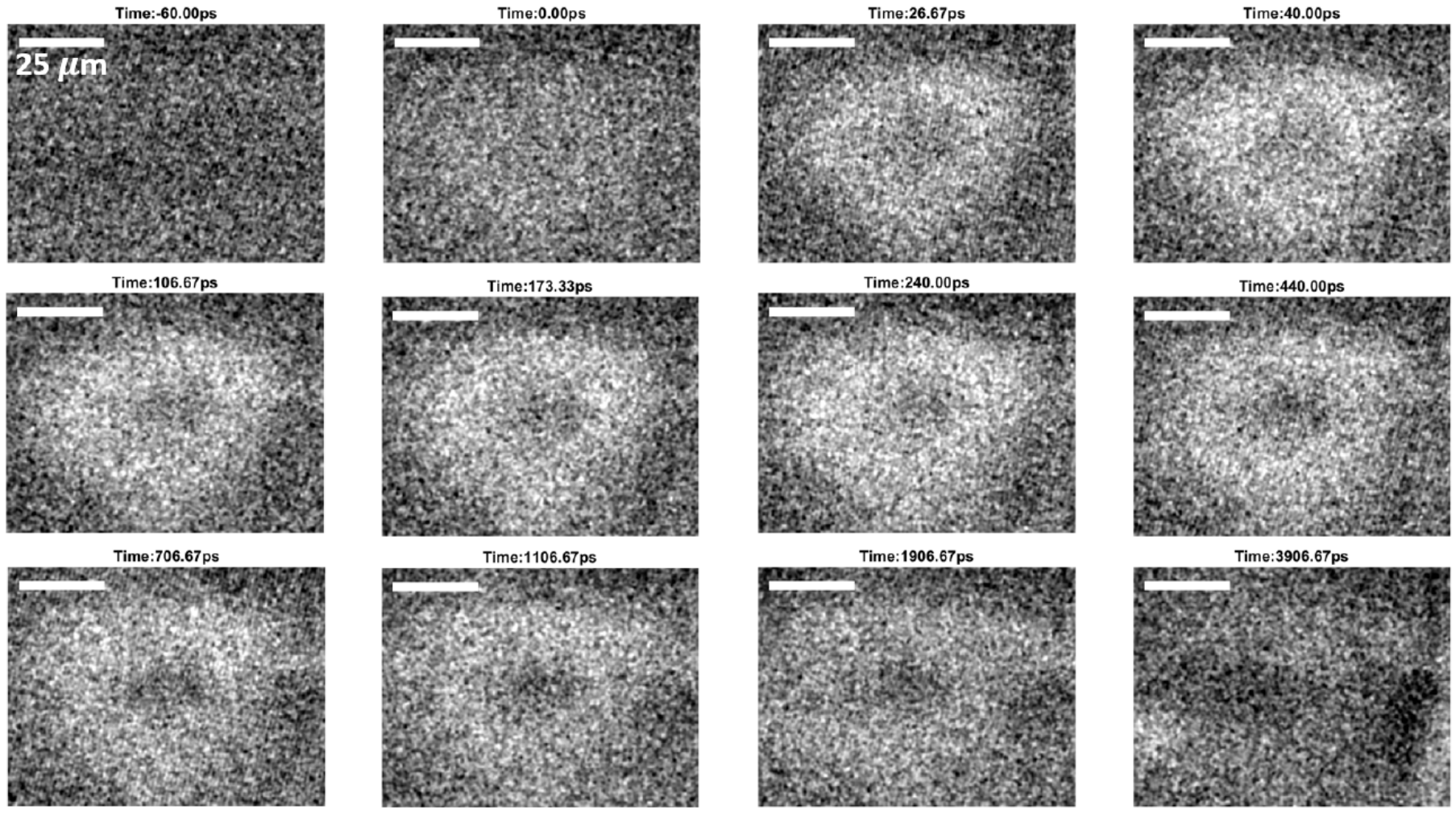

**Fig. S10.** Additional SUEM contrast images taken on $\alpha$-$RuCl_3$. Optical fluence: 250 μJ/cm$^2$.

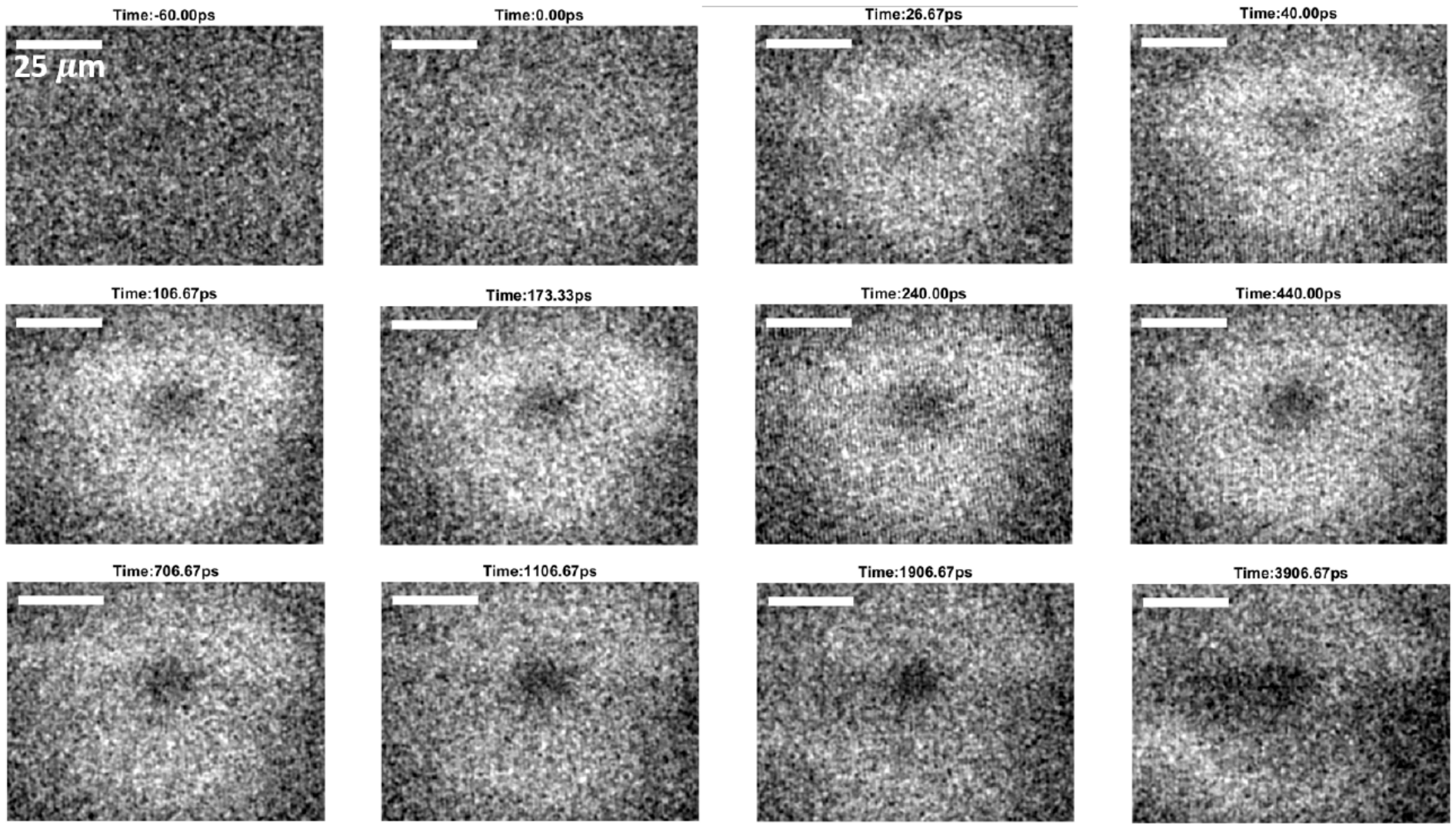

**Fig. S11.** Additional SUEM contrast images taken on $\alpha$-$RuCl_3$**.** Optical fluence: 300 μJ/cm$^2$.

## Supplementary References